\documentclass[12pt]{article}

\usepackage{amsfonts}
\usepackage{epsfig}
\usepackage{a4}
\usepackage{theorem}
\usepackage{amssymb}
\usepackage{graphicx}
\input axodraw.sty

\newcommand{\be}[1]{\begin{equation}\label{#1}}
\newcommand{\ee}{\end{equation}}
\newcommand{\ba}[1]{\begin{eqnarray}\label{#1}}
\newcommand{\ea}{\end{eqnarray}}
\begin{document}
\title{
TWO-LOOP GHOST-ANTIGHOST CONDENSATION FOR $SU(2)$ YANG-MILLS
THEORIES IN THE MAXIMAL ABELIAN GAUGE}
\author{
A.R.~Fazio
\vspace{3mm} \\
\small Bogoliubov Laboratory of Theoretical Physics,
Joint Institute for Nuclear Research,\\
\small Dubna 141980, Russian Federation\\
\small raffaele@thsun1.jinr.ru\vspace{3mm}\\}

\date{} \maketitle
\begin{abstract}
In the framework of the formalism of Cornwall et all. for
composite operators I study the ghost-antighost condensation in
$SU(2)$ Yang-Mills theories quantized in the Maximal Abelian Gauge
and I derive analytically a condensating effective potential at
two ghost loops. I find that in this approximation the one loop
pairing ghost-antighost is not destroyed but no mass is generated
if the ansatz for the propagator suggested by the tree level
Hubbard-Stratonovich transformations is used.

\end{abstract}
\section{Introduction}
The ghost-antighost condensation in $SU(2)$ Yang-Mills theories
quantized in the Maximal Abelian Gauge \cite{Kronfeld} was
proposed by Martin Schaden \cite{schad} in 1999. The original aim
was to investigate how to preserve the methods of perturbation
theory when infrared divergences plague the high temperature phase
of QCD \cite{Kapusta}. In fact, the analysis of Schaden provided
analytical propagators for all fields except for the Abelian
photon due to a dynamically generated screening mass. Later
\cite{KondDudal,SAWA} this phenomenon was connected with a
possible explanation of the Abelian Dominance in non Abelian gauge
field theories.

This analysis was given in the mean field approximation at one
loop order. In this note I will extend this analysis at two-loop
order within the functional formalism of
Cornwall-Jackiw-Tomboulis, which was already used to study the
dynamical mass generation in the model of Cornwall-Norton
\cite{CJT} and the chiral symmetry breaking in quantum
chromodynamics \cite{CasalCasto}. The aim of this note is to shed
light on the dynamics of the ghost condensation. I will prove that
the ghost-antighost propagator suggested at tree level
\cite{schad,KondDudal,SAWA}, using Hubbard-Stratonovich
transformations, is not compatible at quantum level with a
dynamical mass generation.
\section{$SU(2)$ Yang-Mills theories in the Maximal Abelian Gauge}

I shall consider the Maximal Abelian Gauge fixed $SU(2)$
Yang-Mills action in the four dimensional continuum Minkowski
space \cite{MPL}
\begin{eqnarray}
 \textsl{S} &=& \int d^4x \Bigg[-\frac 1{4g^2} F_{\mu\nu}^a F^{a\mu\nu}-\frac 1{4g^2}F_{\mu\nu }F^{\mu\nu }
 -{\frac 1 {2\alpha}}\left(D_\mu ^{ab} A^{b\mu}\right)^2+\nonumber\\
& & +\overline{c}^a D_\mu ^{ab}D^{\mu
bc}c^c-\varepsilon^{ab}\varepsilon^{cd}\overline{c}^a c^d A_\mu ^b
A^{c\mu}-{\frac \alpha 4}\epsilon^{ab}\epsilon^{cd}\bar{c}^a
\bar{c}^b c^c c^d \Bigg]\label{prima}.
\end{eqnarray}
According to \cite{gerardo} I have chosen the diagonal generator
of the gauge group $SU(2)$ as Abelian charge and I have made the
following decompositions for the gluons, ghosts and antighosts
fields respectively: $(A^{\mu a},A^\mu)$, $(c^a,c)$,
$(\bar{c}^a,\bar{c})$, $a=1,2$ labels the off-diagonal components
of the Lie-algebra valued fields.

The covariant derivative $D_\mu ^{ab}$ is defined with respect to
the diagonal component $ A_\mu$ of the Lie Algebra valued
connection
\begin{equation}
D_\mu ^{ab}\equiv \partial _\mu \delta ^{ab}-\epsilon ^{ab}A_\mu.
\label{cder}
\end{equation}
The components of the field strength are:
\begin{eqnarray}
F_{\mu\nu}^a&=&D_\mu^{ab}A_\nu^b-D_\nu^{ab}A_\mu^b\nonumber\\
F_{\mu\nu}&=&\partial_\mu A_\nu-\partial_\nu
A_\mu+\epsilon^{ab}A_\mu^a A_\nu^b.
\end{eqnarray}
In the action $\textsl{S}$ it has been used the partial gauge
fixing condition
\begin{equation}
D_\mu^{ab}A^{\mu b}=0. \label{MAG}
\end{equation}
The action $\textsl{S}$ manifests a residual $U(1)$ gauge symmetry
which can be fixed imposing for example the Landau condition
\begin{equation}
\partial_\mu A^\mu=0.
\label{land}
\end{equation}
In the following I will not consider the Faddeev-Popov terms
related to (\ref{land}) since they don't play any role.

In (\ref{prima}) the value of the gauge parameter $\alpha$ has
been taken equal to the ''coupling constant" of the quartic
ghost-antighost interaction. In the Maximal Abelian Gauge this
interaction is needed for renormalizability and appears at tree
level with an arbitrary coupling in order to remove the
logarithmic divergence of the full two $A_\mu$ and two $A_\mu^a$
exchange between a pair of ghost-antighost scattering \cite{MPL}.
This phenomenon reminds the renormalizability of scalar quantum
electrodynamics \cite{CW}. In particular the model (\ref{prima})
depends on only one parameter, the $U(1)$ invariance is preserved
at every order in perturbation theory as a consequence of the
global symmetry\cite{fazio}
\begin{equation}
c\rightarrow c + \theta
\end{equation}
which allows for the $c$ independence of $\textsl{S}$.

\section{The effective potential and one-loop calculations}
In order to investigate about the dynamical generation of the
condensate
\begin{equation}
<0\mid\bar{c}^a\epsilon^{ab}c^b\mid0>
\end{equation}
I will construct the Hartree-Fock approximation to the generalized
effective potential \cite{CJT} for the model of the previous
section. This effective potential will depend only on the complete
propagators of the theory $G(x,y)$ for the off-diagonal ghosts,
$\Delta_a(x,y)$ and $\Delta(x,y)$ for off-diagonal and diagonal
gluons respectively. A field dependence is not included, since we
do not expect that any of the fields acquire a vacuum expectation
value. Thus for our problem we have:
\begin{eqnarray}
V(G,\Delta_a,\Delta)&=& -\imath\int
\frac{d^4p}{(2\pi)^4}{\rm{tr}}\left[\log(S^{-1}(p)G(p))-S^{-1}(p)G(p)
+1\right]\nonumber\\
&+&\frac{\imath}{2}\int
\frac{d^4p}{(2\pi)^4}{\rm{tr}}\left[\log({D_a}^{-1}(p)\Delta_a(p))-{D_a}^{-1}(p)\Delta_a(p)+1\right]\label{potential}
\\&+&\frac{\imath}{2}\int
\frac{d^4p}{(2\pi)^4}{\rm{tr}}\left[\log({D}^{-1}(p)\Delta(p))-{D}^{-1}(p)\Delta(p)+1\right]
+V_2(G,\Delta_a,\Delta)\nonumber
\end{eqnarray}
In the previous formula all space-time and gauge indices have been
suppressed. $S(p)$, $D_a(p)$ and $D(p)$ are the free propagators:
\begin{eqnarray}
(D_a)^{ab}_{\mu\nu}(p)&=&-\imath\frac{g^2}{p^2}\delta^{ab}
\left[\eta_{\mu\nu}-\frac{(1-\alpha)p_\mu p_\nu}{p^2}\right],\nonumber\\
D_{\mu\nu}(p)&=& -\imath\frac{g^2}{p^2}
\left[\eta_{\mu\nu}-\frac{p_\mu p_\nu}{p^2}\right],\label{props}\\
S^{ab}(p)&=& -\frac{\imath}{p^2}\delta^{ab}.\nonumber
\end{eqnarray}
In order to focus on the ghost-antighost condensation let us
consider the approximation in which $\Delta_a(p)=D_a(p)$ and
$\Delta(p)=D(p)$. It will be proved in the following that the
accuracy of this approximation is under control because I work in
the weak coupling regime, $g^2 << 1$. In this approximation $V_2$
includes the contribution of diagrams which are two-particle
irreducible with respect to ghost-antighost lines only.

To compute the effective potential (\ref{potential}) I make the
following ansatz for the ghost propagator:
\begin{equation}
G^{ab}(p)= -\imath \frac{p^2\delta^{ab}+\varphi
(p^2)\epsilon^{ab}}{p^4+\varphi^2 (p^2)} \label{ghost}
\end{equation}
by defining
\begin{equation}
-\imath\varphi(p^2)\epsilon_{ab}= G^{-1}_{ab}-S^{-1}_{ab}.
\label{conden}
\end{equation}
If $\varphi(p^2)$ is constant the ansatz (\ref{ghost}) agrees with
the ghost propagator used in \cite{schad,KondDudal,SAWA} by making
Hubbard-Stratonovich transformations.

The behaviour of $\varphi(p^2)$ can be seen from the
Dyson-Schwinger equation for the ghost propagator or equivalently
from the mass gap equation \cite{CJT} of the effective potential
$V$. I will investigate about the complete system of the
Dyson-Schwinger equations in the Maximal Abelian gauge in a
subsequent paper. Concerning now I observe that disregarding the
tadpole terms and replacing the complete $A^\mu \bar{c}c$ vertex
by the bare one (Hartree-Fock approximation):
\begin{equation}
\partial_\mu\bar{c^a}\epsilon^{ab}A^\mu
c^b-\bar{c^a}\epsilon^{ab}A^\mu\partial_\mu c^b
\end{equation}
the two-loop part of $V$ is
\begin{equation}
V_2 =-\imath\int \frac{d^4p
d^4q}{(2\pi)^8}\left[p_\rho(p_\mu+q_\mu)G^{fa}(p)\epsilon^{ac}G^{cd}(q)
\epsilon^{df}D^{\mu\rho}(p-q)\right].
\end{equation}
The mass gap equation for (\ref{potential}):
\begin{equation}
\frac{\delta V}{\delta G}=0
\end{equation}
becomes in this approximation, with the definition (\ref{conden}),
\begin{equation}
\varphi(p^2)=-4\imath g^2\int \frac{d^4 q}{(2\pi)^4}
\frac{\varphi(q^2)}{q^2(q-p)^2}, \label{DS}
\end{equation}
where the propagator for $A_\mu$ in the Feynman-gauge has been
used. Nevertheless for a non trivial $\varphi(p^2)$ the equation
(\ref{DS}) is not compatible with the rest coming from the
symmetric part of (\ref{ghost})
\begin{equation}
0=g^2\int \frac{d^4 q}{(2\pi)^4} \frac{q^4}{q^4+\varphi^2(p^2)}
\frac{1}{(p-q)^2}
\end{equation}
If I ignore this important point the result is no mass generation
due to ghost condensation.

The equation (\ref{DS}) is similar in structure to the equation
for the chirally asymmetric part of the inverse electron
propagator in the Baker-Johnson-Willey approach to electrodynamics
\cite{BJW}. Guided by the work of these authors I ask if there is
a solution to (\ref{DS}) whose asymptotic behaviour is:
\begin{equation}
\varphi(p^2)=\left\{\begin{array}{ll}
\varphi & \mbox{$\mid -p^2\mid\leq \Lambda^2$}\\
\varphi (-\frac{p^2}{\Lambda^2})^{-\varepsilon}& \mbox{$\mid
-p^2\mid \gg \Lambda^2$}
\end{array}
\right.\label{fi}
\end{equation}
in which $\Lambda$ is taken as a fixed massive parameter. Of
course $\varphi(p^2)$ must be a continuous function and one should
specify the transition between the high energy and the low energy
behaviour. However various reasonable transition behaviours make
only a small difference in the numerical coefficient of the final
result of the effective potential \cite{CJT}.

The integral equation (\ref{DS}) is equivalent to the following
differential equation:
\begin{equation}
\frac{d}{dx}\left(x^2\frac{d}{dx}\varphi(x)\right)
=-\frac{4g^2}{16\pi^2}\varphi(x).
\end{equation}
If I put the ansatz (\ref{fi}) I obtain, for $g^2<<1$, the
solution
\begin{equation}
\varepsilon=\frac{4g^2}{16\pi^2} +O(g^2). \label{epsilon}
\end{equation}
Because $\varepsilon$ is small, the ansatz (\ref{fi}) is a good
approximation also in the infrared domain \cite{CJT,CN}.
However in the following $\varepsilon \rightarrow 0$, playing the
role of a regulator, therefore in any gauge I will assume an order
of magnitude given by (\ref{epsilon}).

I would like to stress that $\varphi(p^2)\epsilon_{ab}$ is, in my
notation, the antisymmetric part of the propagator $G$ but
$\varphi$ is $p^2$-independent and plays the role of some suitably
regularized value of $<0\mid\bar{c}^a\epsilon^{ab}c^b\mid0>$.

The one-loop contribution to (\ref{potential}) up to
$\varphi$-independent terms is obtained from (\ref{props}) and
(\ref{ghost}):
\begin{eqnarray}
V_1(\varphi)=
-\imath\int\frac{d^4p}{(2\pi)^4}\left[\log\left(1-\frac{\varphi^2
(p^2)}{p^4+\varphi^2(p^2)}\right) +\frac{2
\varphi^2(p^2)}{p^4+\varphi^2(p^2)}\right]. \label{1loop}
\end{eqnarray}

This expression takes the following form in the Euclidean region:
\begin{eqnarray}
V_1(\varphi)= -\frac{1}{16\pi^2}\int_{0}^{+\infty}dx\,\,
x\left[\log\left(1-\frac{\varphi^2(x)}{x^2+\varphi^2(x)}\right)
+\frac{2 \varphi^2(x)}{x^2+\varphi^2(x)}\right]. \label{1loop}
\end{eqnarray}
The evaluation proceeds by inserting (\ref{fi}) into (\ref{1loop})
and keeping only terms that are proportional to inverse power of
$\varepsilon$ as well of zero-order in $\varepsilon$. In practice
I set $\varepsilon$ to zero everywhere as long as no divergence
arises; if $\varepsilon=0$ is not allowed (\ref{fi}) is used.
Therefore
\begin{eqnarray}
&&V_1 (\varphi)= -\frac{1}{16\pi^2}\int_{0}^{\Lambda^2}dx\,\,
x\left[\log\left(1-\frac{\varphi^2}{x^2+\varphi^2}\right) +\frac{2
\varphi^2}{x^2+\varphi^2}\right]\nonumber\\
&&-\frac{1}{16\pi^2}\int_{\Lambda^2}^{+\infty}dx\,\,
x\left[\log\left(1-\frac{\varphi^2\cdot
(\frac{x}{\Lambda^2})^{-2\varepsilon}}{x^2+\varphi^2\cdot
(\frac{x}{\Lambda^2})^{-2\varepsilon}}\right) +\frac{2
\varphi^2\cdot
(\frac{x}{\Lambda^2})^{-2\varepsilon}}{x^2+\varphi^2\cdot
(\frac{x}{\Lambda^2})^{-2\varepsilon}}\right].
 \label{1loopeuc}
\end{eqnarray}
Performing the Laurent expansion around $\varepsilon=0$ we get the
result:
\begin{equation}
V_1 = \frac{\varphi^2}{32\pi^2}-\frac{\varphi^2}{32\pi^2
\varepsilon}+\frac{\varphi^2}{32\pi^2}\log\left(\frac{\varphi^2}{\Lambda^4}\right)
\label{1looprenorm}
\end{equation}
It agrees with the computed result in the $\overline{\rm MS}$
scheme \cite{KondDudal}.

\section{Contributions of Two-Loop Diagrams}
Now let me consider the two-loop contribution to the effective
potential. I am looking for connected, two-particle irreducible
graphs of order $\hbar^2$ in the expression:
\begin{eqnarray}
\imath \hbar<0\mid &&T\exp\left\{-\imath\hbar\int
d^4x\left[\partial_\mu\bar{c^a}\epsilon^{ab}A^\mu
c^b-\bar{c^a}\epsilon^{ab}A^\mu\partial_\mu
c^b+\frac{\alpha}{2}\epsilon^{ab}\epsilon^{cd}\bar{c}^a
c^b\bar{c}^cc^d\right.\right.\nonumber\\&&\left.\left.-\bar{c}^a
c^a A_\mu A^\mu-\epsilon^{ab}\epsilon^{cd}\bar{c}^a c^d A_\mu ^b
A^{\mu c}\right]\right\}\mid 0>, \label{Tprod}
\end{eqnarray}
the parameter $\hbar$ has been introduced in order to count loops,
but it will be put equal to one at the end of the calculation.
Upon scaling the fields in (\ref{Tprod}) like $\psi\rightarrow
\hbar^{1/2}\psi$, expanding the exponential to the relevant order
and applying Wick's theorem, I am left with four integrals. Let me
consider the first one:
\begin{eqnarray}
\frac{I_1}{\hbar^2}&=& g^2\int\frac{d^4 p}{(2\pi)^4}\frac{d^4
q}{(2\pi)^4}\left\{\frac{p_\rho
(p_\mu+q_\mu)}{(p-q)^2}\left(\eta^{\mu\rho}-\frac{(p^\mu-q^\mu)(p^\rho-q^\rho)}
{(p-q)^2}\right)\right.\nonumber\\&&\left.\left[-\frac{2p^2q^2}{(p^4+\varphi^2(p^2))
(q^4+\varphi^2(q^2))}+\frac{2\varphi(p^2)\varphi(q^2)}{(p^4+\varphi^2(p^2))
(q^4+\varphi^2(q^2))}\right] \right\}.
\end{eqnarray}
After making some standard integration on the angles
\cite{Miransky}, I get in the Euclidean region
\begin{eqnarray}
\frac{I_1}{\hbar^2}&=& \frac{3g^2}{256\pi^4}\int_{0}^{+\infty}dx
dy \left\{\frac{x y}{(x^2+\varphi^2(x))(y^2+\varphi^2(y))}
-\frac{\varphi(x)\varphi(y)}{(x^2
+\varphi^2(x))(y^2+\varphi^2(y))}\right\}\times\nonumber\\
&&\left.\times[y^2\theta(x-y)+x^2\theta(y-x)\right].
\end{eqnarray}
Using the expression given in (\ref{fi}) I obtain the following
decomposition
\begin{eqnarray}
\frac{I_1}{\hbar^2}=\frac{3g^2}{128\pi^4}&\times&\left
[\int_{0}^{\Lambda^2}dy
\frac{y^3}{y^2+\varphi^2}\int_{y}^{\Lambda^2}dx
\frac{x}{x^2+\varphi^2}-\int_{0}^{\Lambda^2}dy\frac{y^2}{y^2+\varphi^2}
\int_{y}^{\Lambda^2}dx
\frac{\varphi^2}{x^2+\varphi^2}\right.\nonumber
\\
&+&\left.\int_{0}^{\Lambda^2}dy
\frac{y^3}{y^2+\varphi^2}\int_{y}^{\Lambda^2}dx
\frac{x}{x^2+\varphi^2
\left(\frac{x}{\Lambda^2}\right)^{-2\varepsilon}}\right.\nonumber\\&-&
\left.\int_{0}^{\Lambda^2}dy
\frac{y^2}{y^2+\varphi^2}\int_{\Lambda^2}^{+\infty}dx
\frac{\varphi^2\left(\frac{x}{\Lambda^2}\right)^{-\varepsilon}}
{x^2+\varphi^2\left(\frac{x}{\Lambda^2}\right)^{-2\varepsilon}}
\right.\nonumber\\
&+& \left.\int_{\Lambda^2}^{+\infty}dy
\frac{y^3}{y^2+\varphi^2\left(\frac{y}{\Lambda^2}\right)
^{-2\varepsilon}}\int_{y}^{+\infty}dx \frac{x}{x^2+\varphi^2
\left(\frac{x}{\Lambda^2}\right)^{-2\varepsilon}}\right.\nonumber
\\&-&\left.
\int_{\Lambda^2}^{+\infty}dy\frac{y^2
\left(\frac{y}{\Lambda^2}\right)^{-\varepsilon}}{y^2+\varphi^2\left
(\frac{y}{\Lambda^2}\right)^{-2\varepsilon}} \int_{y}^{+\infty}dx
\frac{\varphi^2\left(\frac{x}{\Lambda^2}\right)^{-\varepsilon}}
{x^2+\varphi^2\left(\frac{x}{\Lambda^2}\right)
^{-2\varepsilon}}\right]. \label{2loop1}
\end{eqnarray}
After making analytical continuation \cite{Vladimir} and Laurent
expansion of (\ref{2loop1}) around $\varepsilon=0$ I get
\cite{grad}:
\begin{equation}
\frac{I_1}{\hbar^2}= \frac{3g^2}{512 \pi^4 \varepsilon}\varphi^2
+\frac{3g^2\varphi^2}{256\pi^4}\left(-\frac{\pi^2}{6}+\frac{1}{2}\right)
-\frac{3g^2}{512\pi^4}\varphi^2\log
\left(\frac{\varphi^2}{\Lambda^4}\right).
\end{equation}
In the appendix I will give more details about how I performed the
integrals of (\ref{2loop1}). Now let me consider the second
integral coming from the expansion of (\ref{Tprod}):
\begin{equation}
\frac{I_2}{\hbar^2}=-\alpha\left\{\left[\int\frac{d^4p}{(2\pi)^4}
\frac{\varphi(p^2)}{p^4+\varphi^2(p^2)}\right]^2+\left[\int\frac{d^4p}{(2\pi)^4}
\frac{p^2}{p^4+\varphi^2(p^2)}\right]^2\right\}.
\end{equation}
Substituting the expression (\ref{fi}) in (\ref{Tprod}) I get for
the first term after the usual change of variables $p^0\rightarrow
\imath p^0$, analytical continuation \cite{Vladimir} around
$\varepsilon=0$:
\begin{equation}
\frac{I_2}{\hbar^2}=\frac{\alpha
\varphi^2}{256\pi^4}\left(-\frac{1}{2}
\log\left(\frac{\varphi^2}{\Lambda^4}\right)
+\frac{1}{\varepsilon}\right)^2.
\end{equation}
and for the second term
\begin{equation}
\int\frac{d^4 p}{(2\pi)^4}\frac{p^2}{p^4+\varphi^2 (p^2)}=
O(\varepsilon), \label{zero}
\end{equation}
It will proved in the appendix.

Finally it is easy to see that the sum of the last two integrals
that can be extracted from (\ref{Tprod}) is:
\begin{equation}
\frac{I_3+I_4}{\hbar^2}=2\int\frac{d^4
p}{(2\pi)^4}\frac{p^2}{p^4+\varphi^2(p^2)}\int\frac{d^4
q}{(2\pi)^4}\frac{\alpha}{q^2}
\end{equation}
and it is $O(\varepsilon)$ due to (\ref{zero}).

By using the same method and defining for massive off-diagonal
gluons the following propagator:
\begin{equation}
(\Delta_a)^{ab}_{\mu\nu}(p)=-\imath\frac{g^2}{p^2-M^2(p^2)}\delta^{ab}
\left[\eta_{\mu\nu}-\frac{(1-\alpha)p_\mu
p_\nu}{p^2-M^2(p^2)}\right]
\end{equation}
with
\begin{equation}
M^2(p^2)=\left\{\begin{array}{ll} M^2 & \mbox{$\mid -p^2\mid\leq
\Lambda^2$}\\ M^2 \cdot(-\frac{p^2}{\Lambda^2})^{-2\varepsilon}&
\mbox{$\mid -p^2\mid \gg \Lambda^2$}
\end{array}
\right.
\end{equation}
it is easy to see that the vertex $\bar{c}cAA$ will provide a
$O(\varepsilon)$ contribution to the effective potential $V$,
which must be disregarded for $g^2<<1$. The main point of this
paper is here. If one uses the ansatz (\ref{ghost}), the effective
potential doesn't possess, at the lowest order in the weak
coupling regime, the necessary mixing term between $M$ and
$\varphi$ for the generation of a mass for off-diagonal gluons
related to the ghost-antighost condensate. That is because the
symmetric part of (\ref{ghost}) doesn't satisfy the
Dyson-Schwinger equations.

Moreover it is possible to say that due to (\ref{zero}) the
approximation $\Delta_a=D_a$ is compatible with the weak coupling
regime. Since the propagator $\Delta$ is supposed to coincide with
the normal perturbative solution because no symmetry-breaking
effects are expected, the weak coupling regime controls also the
approximation $\Delta=D$.
\section{Effective potential and the ghost condensate}
Collecting the results found in the previous section and keeping
only terms that are proportional to inverse powers of the coupling
$g$ (these come from inverse powers of $\varepsilon$) as well as
terms of zeroth order in $\varepsilon$ and coupling I get the
two-loop effective potential
\begin{eqnarray}
V(\varphi)&=&
\frac{\varphi^2}{32\pi^2}\left(1-\frac{1}{\varepsilon}\right)
+\frac{1}{32\pi^2}\varphi^2\log
\left(\frac{\varphi^2}{\Lambda^4}\right)\nonumber\\
&+&\frac{\alpha\varphi^2}{256\pi^4}\left(\frac{1}{2}
\log\left(\frac{\varphi^2}{\Lambda^4}\right)-\frac{1}{\varepsilon}
\right)^2, \label{finpot}
\end{eqnarray}
where terms divergent at $\varepsilon=0$ but multiplied by higher
powers of the coupling constant have been dropped.

In the weak coupling regime the effective potential is independent
on the gauge parameter $\zeta$ of the $U(1)$ symmetry. In fact it
is easy to check, using the results of the previous section, that
in a general covariant gauge one should add to (\ref{finpot})
\begin{equation}
\frac{\zeta g^2\varphi^2}{256\pi^4}\left(\frac{1}{2}
\log\left(\frac{\varphi^2}{\Lambda^4}\right)-\frac{1}{\varepsilon}
\right)^2
\end{equation}
which is negligible compared to the term proportional to $\alpha$
if $\alpha >> g^2$.

Although (\ref{finpot}) is not the end of the story, it is worth
to remark that the effective potential $V(\varphi)$ must be
bounded from below therefore:
\begin{equation}
\alpha > 0
\end{equation}
which is equivalent to state the concavity of
$V(\varphi)$\cite{IIM}. Moreover since this potential manifests a
nontrivial absolute minimum if
\begin{equation}
\alpha> -\frac{\varepsilon}{4} \label{ineq}
\end{equation}
and since we work for $\varepsilon\rightarrow 0$ the inequality
(\ref{ineq}) is satisfied if $V(\varphi)$ is bounded from
below.The absolute minimum of our effective potential is found to
be at:
\begin{equation}
\log\left(\frac{\varphi^2}{\Lambda^4}\right)=
\frac{2}{\varepsilon}-1-\frac{16\pi^2}{\alpha}.
\end{equation}

I observe that the quartic ghost-antighost interaction seems to
play a crucial role in this mechanism of condensation. This
interaction seems to affect the effective potential much more than
the cubic vertex $\bar{c}cA$ which only perturbs the one-loop
result. It is worth to remark that $\alpha$ positive could be
related to a sort of "ghost-attraction", but unfortunately I don't
have any general argument to state the positivity of
\begin{equation}
-{\frac \alpha 4}\epsilon^{ab}\epsilon^{cd}\bar{c}^a \bar{c}^b c^c
c^d = \alpha\bar{c}^1c^1\bar{c}^2c^2
\end{equation}
when the usual assignments of hermiticity \cite{KO}
\begin{eqnarray}
c^\dagger &=& c\nonumber\\
\bar{c}^\dagger &=& -\bar{c}
\end{eqnarray}
are given.

The contributions to the effective potential proportional to
$\alpha$ are dominated by the term
\begin{equation}
\frac{\alpha
\varphi^2}{1024\pi^4}\log^2\left(\frac{\varphi^2}{\Lambda^4}\right)
\end{equation}
which is clearly a symmetry restoring term. Nevertheless if
\begin{equation}
\alpha\thickapprox16\pi^2 \varepsilon
\end{equation}
the absolute minima of $V_1$ and $V$ are on the same value and
easily it is possible to see:
\begin{equation}
\left(\frac{V}{\varphi^2}-\frac{V_1}{\varphi^2}\right)_{\rm min}=
O(g^2).
\end{equation}
Therefore for $\alpha\sim O(g^2)$ the two-loop contribution
corresponds to a small perturbation of the one-loop result.
\section{Discussion}
I have derived a two-loop ghost-antighost condensating effective
potential in the weak coupling regime using an ansatz found at
tree level, but not efficient at quantum level. The consequence of
this wrong ansatz has been that in the off-diagonal gluon
propagator no mass or infrared cut-off is generated as claimed in
\cite{schad,KondDudal}. In order to improve this study it becomes
mandatory to know more about the complete propagators of the
theory, for example from the complicated system of the
Dyson-Schwinger equations. The complete propagators are expected
to show a richer structure than in (\ref{ghost}) due to the
dependence on the most general BRST invariant\cite{KonSlav}
condensate of dimension two
\begin{equation}
<0|A_\mu^a A^{\mu a}+\alpha\bar{c}^a c^a|0>. \label{BRST}
\end{equation}

This condensate for Yang-Mills theories in the Maximal Abelian
Gauge is now under investigations \cite{stronzi}. For its
computation could be crucial the residual $U(1)$ gauge invariance
of the theory after the partial gauge fixing condition
(\ref{MAG}). In fact taking $\alpha=-1$ and calling $\xi$ the
coupling constant of the self interaction between ghosts the
action can be written:
\begin{eqnarray}
S&=&\int d^4x\left[-\frac{1}{4g^2}(\partial_\mu A_\nu-\partial_\nu
A_\mu)^2 +\frac{1}{2g^2}{A_\nu}^a D_\mu ^{ab}{D^\mu}_{bc}A^{\nu
c}-\frac{1}{2g^2}(\epsilon^{ab}{A_\nu}^a A_\mu^b)^2\right.\nonumber\\
&& \left.+\overline{c}^a D_\mu ^{ab}D^{\mu
bc}c^c+\frac{\xi}{2}(\bar{c}^a
c^a)^2-\varepsilon^{ab}\varepsilon^{cd}\overline{c}^a c^d
{A_\mu}^b A^{c\mu}\right].
\end{eqnarray}
This action represents a sort of scalar electrodynamics of charged
off-diagonal gluons and the off-diagonal ghosts and antighosts
fields interacting each other by usual quartic scalar terms. These
classes of models, constraints by the vanishing of the vacuum
expectation value of every charged scalar fields, provided a
stable vacuum due the pair condensates of the charged scalar
fields \cite{Mand}. Using these results the condensate
(\ref{BRST}) could be evaluated providing a gauge invariant mass
generation for continuum Yang-Mills theories.

\appendix
\section{Integrals}

In this appendix I will give more details about the calculations
of two integrals met in section 3.

The first integral
\begin{equation}
J_1=\int_{\Lambda^2}^{+\infty}dy
\frac{y^3}{y^2+\varphi^2\left(\frac{y}{\Lambda^2}\right)
^{-2\varepsilon}}\int_{y}^{+\infty}dx \frac{x}{x^2+\varphi^2
\left(\frac{x}{\Lambda^2}\right)^{-2\varepsilon}}
\end{equation}
is easily shown to be equal to :
\begin{equation}
\Lambda^4\int_{1}^{+\infty}dy \frac{y^3}{y^2+f^2 y
^{-2\varepsilon}}\int_{y}^{+\infty}dx \frac{x}{x^2+f^2
x^{-2\varepsilon}}.
\end{equation}
with $f^2=\frac{\varphi^2}{\Lambda^2}$. Since\cite{grad}
\begin{equation}
\int_{y}^{+\infty}dx \frac{x}{x^2+f^2 x^{-2\varepsilon}}=
\frac{1}{2(1+\varepsilon)}\log\left(1+\frac{y^{2+2\varepsilon}}{f^2}\right)
\end{equation}
our integral becomes in its convergence region \cite{Vladimir}
\begin{equation}
J_1=\frac{\Lambda^4}{2(1+\varepsilon)}\int_{1}^{+\infty}dy
\frac{y^{3+2\varepsilon}}{y^{2+2\varepsilon}+f^2}
\log\left(1+\frac{y^{2+2\varepsilon}}{f^2}\right)
\end{equation}
I adopt the following trick
\begin{eqnarray}
J_1&=&\frac{\Lambda^4}{2(1+\varepsilon)}\int_{1}^{+\infty}dy
\left[\left(\frac{y^{3+2\varepsilon}}{y^{2+2\varepsilon}+f^2}
-y+\frac{f^2}{y}
\right)\log\left(1+\frac{y^{2+2\varepsilon}}{f^2}\right)\right.\nonumber\\
&&\left.+\left(y-\frac{f^2}{y}
\right)\log\left(1+\frac{y^{2+2\varepsilon}}{f^2}\right)\right].
\end{eqnarray}
But
\begin{eqnarray}
&&\int_{1}^{+\infty}dy
\frac{f^4-f^2y^2+f^2y^{2+2\varepsilon}}{y(f^2+y^{2+2\varepsilon})}\log\left(1+\frac{y^{2+2\varepsilon}}{f^2}\right)
=\nonumber\\&&\int_{1}^{+\infty}dy \frac{f^4}{y(f^2+y^{2})}
\log\left(1+\frac{y^2}{f^2}\right)+O(\varepsilon)=\nonumber\\
&&\frac{f^2}{12}\left[3\log^2\left(1+\frac{1}{f^2}\right)+\pi^2+6{\rm
Li}_2\left(-\frac{1}{f^2}\right)\right]+O(\varepsilon)
\end{eqnarray}
where in the last equality ${\rm Li}_2(x)$ is the dilogarithm
function, with the property:
\begin{equation}
{\rm Li}_2(x)+{\rm Li}_2(1-x)=\frac{\pi^2}{6}-\log x\log(1-x).
\end{equation}
and it has been used the change to the variable
$z=\log\left(1+\frac{y}{f^2}\right)$.

Moreover\cite{grad}
\begin{eqnarray}
\int_{1}^{+\infty}dy
&&y\log\left(1+\frac{y^{2+2\varepsilon}}{f^2}\right)=
\nonumber\\&&\frac{f^2}{2\varepsilon}+\frac{1}{2}\left[(1+f^2)
\left(1-\log\left(1+\frac{1}{f^2}\right)\right)-f^2\log
f^2\right]+O(\varepsilon).
\end{eqnarray}
Finally\cite{grad}
\begin{equation}
\int_{1}^{+\infty}\frac{dy}
{y}\log\left(1+\frac{y^{2+2\varepsilon}}{f^2}\right)=
\frac{f^2}{2}{\rm Li}_2\left(-\frac{1}{f^2}\right)+O(\varepsilon).
\end{equation}
The final result is
\begin{eqnarray}
J_1 &=& \frac{\varphi^2}{2\varepsilon}+\frac{\Lambda^4}{2}-
\frac{\Lambda^4}{2}\log\left(1+\frac{1}{f^2}\right)+\frac{\varphi^2}{4}
\log^2\left(1+\frac{1}{f^2}\right)\nonumber\\
&&-\frac{\varphi^2}{2}\log\left(1+\frac{1}{f^2}\right) +\varphi^2
\left(\frac{\pi^2}{12}+\frac{1}{2}\right)+O(\varepsilon)
\end{eqnarray}
Now I will prove the (\ref{zero}):
\begin{equation}
\int\frac{d^4 p}{(2\pi)^4}\frac{p^2}{p^4+\varphi^2 (p^2)}=
O(\varepsilon). \label{appzero}
\end{equation}
Using hyperspherical Euclidean coordinates the integral becomes
proportional to
\begin{eqnarray}
\int_0^{\Lambda^2}dx \frac{x^2}{x^2+\varphi^2}
+\int_{\Lambda^2}^{+\infty}dx \frac{x^2}{x^2+\varphi^2
x^{-2\varepsilon}}.
\end{eqnarray}
But \cite{grad}
\begin{eqnarray}
&&\int_{\Lambda^2}^{+\infty}dx
\frac{x^2}{x^2+\frac{\varphi^2}{\Lambda^2}x^{-2\varepsilon}}=
-\frac{\Lambda^2}{(3+2\varepsilon)\varphi^2}\,\,\,
_{2}F_{1}\left(1,\,\frac{3+2\varepsilon}{1+2\varepsilon},\,
\frac{4+4\varepsilon}{1+2\varepsilon},
\,-\frac{\Lambda^2}{\varphi^2}\right)\label{integral}\\
&&{\rm If}\,\,\,\, Re(\varepsilon)<-\frac{3}{2}\nonumber.
\end{eqnarray}
Since (\ref{integral}) can be prolonged \cite{Vladimir} at
$\varepsilon=0$, the Laurent expansion of (\ref{appzero}) is
there:
\begin{eqnarray}
\Lambda^2\left(-1+\frac{\varphi}{\Lambda^2}\arctan\frac{\Lambda^2}
{\varphi}+ O(\varepsilon)\right)=-\int_0^{\Lambda^2}dx
\frac{x^2}{x^2+\varphi^2}+O(\varepsilon)
\end{eqnarray}
and I get the result (\ref{appzero}).

\section*{Acknowledgments}
I thank Prof. A.A.Slavnov for many fruitful and enjoyable
discussions. I am indebted to the Joint Institute for Nuclear
Research for kind hospitality and particularly I thank Prof. V.V.
Nesterenko and Prof. V.N. Pervushin for useful discussions. This
work is supported by the NATO/CNR Advanced Fellowships Programme
2002, Adv. 215.35.

\end{document}